\documentstyle[aps,epsf,prl]{revtex}

\begin{document}
\draft
\title{Radiative corrections to polarization observables
for elastic $e+A$-scattering. \\
Part I: Virtual Compton Scattering}
\author{M. P. Rekalo}
\address{\it National Science Center KFTI,
310108 Kharkov, Ukraine}
\author{E. Tomasi-Gustafsson}
\address{\it DAPNIA/SPhN, CEA/Saclay, 91191 Gif-sur-Yvette Cedex,
France}
\maketitle

\date{\today}
\begin{abstract}

We calculate polarization phenomena for virtual Compton scattering on protons,
at relatively large momentum transfer 1 GeV$^2$ $\le -q^2\le$ 5 GeV$^2$ on the basis of a model for  $\gamma^*+ p\to \gamma+p$ with two
main contributions: $\pi^0$-exchange in $t-$channel and $\Delta$-excitation in
$s$-channel. This model applies from threshold to $\Delta$ region. The parameters entering in this model, such as coupling constants and electromagnetic form factors are well known. The analyzing powers for $\gamma^*+\vec p\to \gamma+ p$ and the components of the final proton polarization in $\gamma^*+ p\to \gamma+ \vec p$ are large in absolute value and show strong  sensitivity to $\pi\bigotimes\Delta$ interference. These results can be applied to the calculation of radiative corrections to polarization phenomena in elastic $ep$-scattering.

\end{abstract}

\section{Introduction}

A precise experiment
recently performed at JLab,  with longitudinally 
polarized electrons and measurement of the polarization of the outgoing proton
\cite{Jo00}, $\vec e^-+p\to  e^-+\vec p$, showed that the ratio between the electric and the magnetic proton form factors, $G_{Ep}/G_{Mp}$, strongly 
deviates from unity, as the momentum transfer $-q^2$ increases. This means that if one assumes that the magnetic form factor has a dipole behavior,
following  $\mu_pG_D$, where $G_D$ is the dipole form factor: $G_D=1/\left [1-{q^2}/{0.71 ~\mbox{GeV}^2}\right ]^2$, then $G_{Ep}$ decreases faster than $G_D$ and $G_{Mp}$ in the measured interval. In other words, the charge distribution inside the proton differs from the magnetic distribution, contrary to what has been currently assumed. These results evidently arise new theoretical interest to the problem of radiative corrections to polarization observables in elastic electron-proton scattering. It is well known that radiative corrections are particularly important for high resolution experiments. In case of unpolarized particles, radiative corrections to the cross section in elastic $ep$-scattering
are usually included in the experimental data, according the calculations of Mo
and Tsai \cite{Mo69}.
The size of these corrections at large momentum transfer, $q^2$, become very
important, reaching 30-50 $\%$ at $q^2\simeq 3$ GeV$^2$.  On the other side, the effects of radiative corrections on polarization
observables are only recently studied \cite{Ma00,Ma99}. 
In ref. \cite{Ma00} it was shown that, in the limiting case of the soft photon approximation, the matrix element of elastic $ep$-scattering can be factorized. In this limit all polarization phenomena can not depend on radiative corrections, which cancel exactly.  The two-photon mechanism enters only as a compensation of the infrared divergency: the standard calculations of radiative corrections for $eh$ scattering contain
the contribution of two-photon exchange where most of the transferred momentum
is carried by one virtual photon, while the other photon has very small momentum. 

But more than 30 years ago, it was observed \cite{Gu73,Fr73,Bo73,Le73}  that
in the two-photon exchange mechanism, another piece of phase space, corresponding to equivalent 
sharing of momentum transfer between the two virtual photons, 
must be taken into account. And the steep decreasing of the nucleon form factors with $q^2$ results, in this case, in an essential increasing of the relative role of two-photon exchange and in the violation of the simple $\alpha$-counting rule. Semi-quantitative considerations have shown that such contribution must be carefully taken into consideration in the differential cross section and in polarization phenomena as well.

Experimental evidence of such mechanism, through its interference with the main one-photon exchange, has been searched in a model independent re-analysis of
the recent $ed$ elastic scattering data at relatively large momentum transfer \cite{Re99}. 

The diagrams to be considered typically in any analysis of radiative corrections at the lowest order in $\alpha$ are the following (Fig. 1):

\begin{enumerate}
\item the one-photon exchange diagram (Fig. 1a);

\item the two-photon contributions (double virtual Compton scattering on the hadron A) (Figs. 1b and 1c);

\item the Bethe-Heitler mechanism (Figs. 1d and 1e);

\item Virtual Compton scattering (VCS) (Fig. 1f).

\end{enumerate}

One expects, a priori, that the interference between these mechanisms (more exactly between the mechanisms 1 and 2 or 3 and 4) plays an important role. 

Radiative corrections to polarization phenomena in elastic $ep$ scattering, using the standard QED technique (Bethe-Heitler mechanism and infrared box) has been calculated in ref. \cite{Af01}, and gives expected small effects.

Larger effects are predictable in calculations including the above two-photon exchange mechanism. Such calculation, even in model-dependent form, has to be performed. The same is correct also for the VCS contribution, which is the object of the present paper.

\section{Model for $\gamma^*+p\to \gamma +p$}

We will consider here the process $\gamma^*+p\to \gamma +p$ and calculate the
contributions to the cross section and to polarization observables due to  
the two mechanisms described by the Feynmann diagrams shown in Fig. \ref{fig:fig2}. We are interested in a specific kinematical 
region, concerning the effective mass of the final $\gamma +p$-system: from threshold to the $\Delta$-region at relatively large momentum transfer, $q^2$, in the space-like region, more precisely, 
for 1 GeV$^2\le -q^2\le $ 5 GeV$^2$. This region is particularly interesting in
view of the interpretation of recent data on elastic scattering of 
longitudinally polarized electrons by protons, with measurement of the polarization of the scattered proton \cite{Jo00}. The information on electromanetic form factors that can be extracted is very important for hadron electrodynamics, therefore the estimation of the radiative corrections for polarization observables in elastic $ep$-scattering appears a necessary procedure. Virtual Compton scattering,
$\gamma^*+p\to \gamma +p$, is an important ingredient of the complete program for the calculation of radiative corrections for $ep$-elastic scattering described above. 

Moreover VCS, in these specific kinematical conditions, (i.e. large momentum transfer $q^2$ in space-like region, and at threshold values of  the invariant total energy of the produced $\gamma +p$ system, $W$) has also, by itself, 
an independent physical meaning \footnote{For a recent review on this subject, see \protect\cite{Gu98}}. 

The VCS process that we consider here is a large extrapolation in $q^2$ from the real Compton scattering (with $q^2=0$, but in the same region of $W$). Therefore it can be interesting for the understanding of the 'deformation' of the relative role of two possible mechanisms for VCS, the one-pion exchange (in the $t-$channel) and the $\Delta$-excitation (in the $s-$channel). Both corresponding electromagnetic form factors, namely $F_{\pi\gamma^*\gamma}(q^2)$ (for the vertex $\gamma^*\to \pi\gamma$) and the magnetic form factor $G(q^2)$ (for the vertex $\gamma^* N\to \Delta$) are experimentally well known, in the space-like region, up to relatively large momentum transfer square. But the theoretical understanding of the $q^2$ behavior of these form factors lies on a different level. The $q^2$ behavior of $F_{\pi\gamma^*\gamma}(q^2)$, in this $q^2$ region, has been well understood in terms of perturbative QCD \cite{qcd}: the good agreement between theory and experiment concerning this form factor is typically considered as one of the best successes of perturbative QCD. 

On the contrary, the situation with the magnetic form factor for the transition $N\to\Delta$ differs in many respects: the $q^2$ behavior deviates from the 'standard' dipole dependence; the helicity conservation for the vertex $\gamma^* N\to \Delta$ at  large momentum transfer is strongly violated. Recent JLab data \cite{Fr99} on the process 
$e^- +p\to e^- +\Delta$ clearly show that, up to $-q^2=4$ GeV$^2$, the form factor of the M1-transition is still strongly dominating over the quadrupole (transversal and longitudinal) form factors, whereas  helicity conservation predicts \cite{Br95} that the magnetic and electric quadrupole form factors must be comparable (at large $-q^2$).

Note also that the $q^2$ behavior of the mesonic $F_{\pi\gamma^*\gamma}(q^2)$
and the baryonic $G(q^2)$ form factors are very different: the form factor for the 
$\gamma^*\pi \gamma$-vertex has essentially a monopole dependence, $F_{\pi\gamma^*\gamma}(q^2)\simeq 1/q^2$ (for $-q^2\ge $ 2 GeV$^2$), whereas 
$G(q^2)$ is characterized by a more steep $q^2$-dependence, i.e.:
$$G(q^2)=\displaystyle\frac{G_D(q^2)}{\left (1-\displaystyle\frac{q^2}{m_x^2}\right )},~\mbox{~with~}m_x^2\simeq 6 \mbox{~GeV}^2.$$
Therefore, one can expect that the relative role of $\pi$-exchange increases with increasing $-q^2$.

Another interesting aspect of the suggested model (based on 
$\pi+\Delta$-exchange) for $\gamma^*+p\to \gamma +p$ is the complexity of the resulting amplitude, due to the $\Delta$-contribution, which is at the origin 
of rich polarization phenomena.

T-even polarization observables in VCS, induced by collisions of circularly polarized virtual photons with polarized proton target 
($\vec\gamma^*+\vec p\to \gamma +p$) as well as induced by collisions of circularly polarized virtual photons with unpolarized proton target, but with measurement of polarization of the outgoing proton ($\vec\gamma^*+p\to \gamma +\vec p$) have to be calculated (in the framework of a model) as a necessary part of the program for the calculation of radiative corrections to the processes $\vec e^-+\vec p\to e^-+p$ or 
$\vec e^-+ p\to e^-+ \vec p$.  

Evidently the longitudinally polarized electrons interact with protons through circularly polarized virtual photons.

The matrix element for $\pi^0-$exchange can be written in the following form:
\begin{equation}
{\cal M}_{\pi}=ig\displaystyle\frac{e^2}{t-m_{\pi}^2}
\displaystyle\frac{F_{\pi\gamma^*\gamma}(q^2)}
{m_{\pi}}\epsilon_{\mu\nu\alpha\beta}\epsilon_{\mu}
q_{\nu}e^*_{\alpha}k_{\beta} \overline{u(p_2)}\gamma_5u(p_1),
\end{equation}
where $t=(k-q)^2$, $\epsilon$ and $q$ ($e$ and $k$) are the 4-vector of polarization and the 4-momentum of virtual (real) photon in 
$\gamma^*+p\to \gamma +p$, $g$ is the coupling constant of $\pi^0pp$-vertex,
$F_{\pi\gamma^*\gamma}(q^2)$ is the form factor of the $\pi^0\gamma^*\gamma$-vertex with one virtual photon, $m_{\pi}$ is the pion mass. Following the vector dominance model (VDM), the form factor  
$F_{\pi\gamma^*\gamma}(q^2)$ can be parametrized in a simple way:
$$ F_{\pi\gamma^*\gamma}(q^2)=\displaystyle\frac{F_{\pi}}{1-\frac{q^2}{m_V^2}},$$
where ${F_{\pi}}$ characterizes the decay $\pi^0\to\gamma\gamma$. A similar approach is working also for the heavy $\eta_c$-meson \cite{EdS00}. The parameter 
$m_V\simeq 0.8$ GeV, is comparable to the mass of $\rho$- and $\omega$-mesons.
The constant ${F_{\pi}}$ can be estimated from the width of $\pi^0\to\gamma\gamma$:
$$\Gamma(\pi^0\to 2\gamma)=\displaystyle\frac{\alpha^2}{4}\pi F_{\pi}^2
m_{\pi}.$$
Using the value $\tau(\pi^0)=8.4\cdot 10^{-17}$ s \cite{pdg},
one can find the following value: $F_{\pi}^2= 0.0140$.
The absolute sign of this constant can not be fixed, in this way.
In principle this sign can be determined by the quark model or by QCD-considerations.

The most convenient way to write the matrix element ${\cal M}_{\Delta}$ (for $\Delta$-excitation in s-channel) can be realized in the CMS of 
$\gamma^*+p\to \gamma +p$, using a 2-component formalism:
$${\cal M}_{\Delta}=\displaystyle\frac{e^2\mu^2(\Delta\to N)}{(M+m)^2}
\sqrt{(E_1+m)(E_2+m)}\chi_2^\dagger
(\delta_{ab}-\frac{i}{2}\sigma_{abc}\sigma_c)\chi_1$$
\begin{equation}
\times\displaystyle\frac{2}{3}
\displaystyle\frac{2M}{W^2-M^2+iM\Gamma}
(\vec \epsilon\times\vec q)_a~(\vec e^*\times\vec k)_b~G(q^2),
\label{eq:eq2}
\end{equation}
where $M(m)$ is the mass of $\Delta (N)$, $E_1$ and $E_2$ are the energies of the initial and final protons in $\gamma^*+p\to \gamma +p$, which are described by the 2-component spinors $\chi_1$ and $\chi_2$, $\vec k$ and $\vec q$ are the 3-momenta of the real and virtual photon, $\mu(\Delta\to N)$ is the magnetic moment of the radiative decay $\Delta \to N +\gamma$, $\Gamma$ is the total width of the $\Delta$-isobar. 

The presentation of the  matrix element 
${\cal M}_{\Delta}$ in the form (\ref{eq:eq2}) has the following advantages, in comparison with its relativistic invariant, i.e. the Feynman formulation:
\begin{itemize}
\item more simple and transparent evidence of the M1 nature of the two possible electromagnetic vertexes in the $s-$channel diagram, for $\gamma^*+p\to \gamma +p$;
\item essential simplification of the resulting matrix element;
\item avoiding problems with different off-mass shell effects, which enter in any relativistic description of Feynman diagrams with fermionic propagators, such as contributions of states with different values of spin and parity (virtual antibaryon contribution);
\item the gauge invariance of ${\cal M}_{\Delta}$ contribution is automatically solved from both sides: for real photon (with polarization $\vec e$) and for the virtual photon (with polarization $\vec \epsilon$);
\item each part of the matrix element ${\cal M}_{\Delta}$ has evident physical interpretation.
\end{itemize}
Note that the matrix element ${\cal M}_{\Delta}$ describes the single multipole transition for $\gamma^*+p\to \gamma +p$, namely $\gamma^*(M1)+p\to \gamma (M1)+p$, which dominates in the ${\Delta}-$region. The quadrupole vertexes for real and virtual photons in $\gamma^*+p\to {\Delta}$ and ${\Delta}\to N+\gamma $ are negligible up to large momentum transfer \cite{Fr99}.

Let us write the expressions which connect the radiation width 
${\Delta}\to N+\gamma$ with the transition magnetic moment $\mu(\Delta \to N)$.
The starting point is the following formula for the matrix element which holds in the rest frame of $\Delta$:
$${\cal M}({\Delta}\to N+\gamma)=\displaystyle\frac{e\mu(\Delta\to N)}{M+m}
\chi^\dagger \vec\phi\cdot\vec e\times\vec k \sqrt{2M(E+m)},
$$
where $E$ is the energy of the final $N$ in $\Delta\to N+\gamma$.
One finds:
$$ \Gamma({\Delta}\to N+\gamma) = 
\displaystyle\frac{\alpha}{24}\mu^2(\Delta\to N)M\left (1-\displaystyle\frac{m^2}{M^2}\right )^3.
$$
Using the existing data from \cite{pdg}, concerning $\Delta\to N+\gamma$, one finds: $ \mu^2(\Delta\to N+\gamma)\simeq  22$. 

Summarizing this discussion, we can say that, in the framework of the suggested model, all important ingredients such as electromagnetic and strong coupling constants are fixed, as well as the $q^2$ behavior of both electromagnetic form factors for the vertexes $\pi\gamma^*\gamma$ and $\gamma^*N\to \Delta$ (in the space-like region). However we can do the following remarks, about this model:
\begin{itemize}
\item In principle both diagrams ($\pi$- and $\Delta$-exchange) must contain  phenomenological form factors: $F(t)$- for the $t$-channel and $F(s)$- for the $s$-channel. These form factors are not well known, and their presence may play a role in the preliminary estimation of the polarization phenomena for VCS, which is the primary aim of this work.
One has to take care of the fact that the introduction of such form factors, (in any form) can not violate the electromagnetic current conservation for $\gamma^*+p\to \gamma +p$, with respect to the real and virtual photon, as well.
\item The combination ${\cal M}_{\pi}+{\cal M}_{\Delta}$, as it was mentioned before, results in a complex matrix element, so the polarization phenomena in 
$\gamma^*+p\to \gamma +p$ (the T-even and the T-odd as well) are, in principle, sizeable.
\item In the considered specific case of the process $\gamma^*+p\to \gamma +p$, which is of the second order in the electromagnetic constant (i.e. without strong interaction in the initial and final states), the delicate problem of the possible violation of Christ and Lee theorem is absent \cite {Ch66}. So with any relative phase phase of the matrix elements ${\cal M}_{\pi}$ and ${\cal M}_{\Delta}$, we can avoid the problem of the 'artificial' violation of the T-invariance of hadronic electromagnetic interaction, which is typically present in the discussion of processes of photo- and electroproduction of mesons on nucleons.

\end{itemize}

Clearly the present model is well adapted also for the VCS on a neutron target, 
$\gamma^*+n\to \gamma +n$, with matrix element:
$${\cal M}(\gamma^* n\to \gamma n)= -{\cal M}_{\pi}+{\cal M}_{\Delta}.$$
So, the simultaneous study of both reactions, $\gamma^*+p\to \gamma +p$ and $\gamma^*+n\to \gamma +n$ in principle can allow to solve an old debate concerning the absolute sign of the product of two constants, $gF_{\pi}(0)$,
which appeared in the literature concerning Compton scattering 40 years ago. In the present calculation we will consider both relative signs for ${\pi}$- and ${\Delta}$-contributions. This will also be indicative of the sensitivity of the observables to the ${\cal M}_{\pi}\bigotimes {\cal M}_{\Delta}$-interference phenomenum.
\section{Exclusive cross section for $p(e,e'\gamma) p$}
The differential cross section for the exclusive process 
$\vec e^-+p\to e^-+p+\gamma$, with detection of the final electron and 
proton can be written in the following form, which holds for the one-photon mechanism:
$$
\frac{d^3\sigma }{dE_2 d\Omega_e d\Omega} =\frac{\alpha ^2}
{64\pi ^3}\frac{\epsilon_2}{\epsilon_1}\displaystyle\frac{\omega }{mW}\displaystyle\frac
1{1-\epsilon }\displaystyle\frac{1}{( -q^2) }(E_1+m)(E_2+m),
$$ 
$$X=X_0+\lambda X_1,$$
where $\epsilon_1$ ($\epsilon_2$) is the energy of the initial (final) electron in the LAB system, $\omega$ is the real photon energy in the CMS of $\gamma^*+p\to p+\gamma $, $d\Omega_e $ ($d\Omega$) is the element of solid angle of the final electron (produced real photon) in LAB (CMS), $\epsilon$ is the polarization of the virtual photon, $\epsilon^{-1}=1-2
\displaystyle\frac{\vec {q^2}}{q^2}\tan^2\displaystyle\frac{\theta_e}{2},$ and $\theta_e $ is the electron scattering angle in LAB of the reaction $\vec e^-+p\to e^-+p+\gamma$.
The combination $X=X_0+\lambda X_1$ describes the dedependence of the
differential cross section on the longitudinal polarization $\lambda $ of the
initial electron (so the helicity $\lambda_e$ of the electron can be determined
as $\lambda_e=\lambda/2$) and on the polarization $\vec {\cal T}$  of the proton target.

Let us consider unpolarized collisions, where the $X_0$ can be written in the standard form, in terms of four structure functions:
$$X_0=W_1+\epsilon \cos 2\phi W_2-2\epsilon \displaystyle\frac{q^2}
{\vec {q^2}}W_3-\cos \phi\sqrt{2\epsilon (1+\epsilon)\displaystyle\frac{( -q^2)}
{\vec {q^2}}}W_4,
$$
where $\phi$ is the azymuthal angle (between the plane of electron scattering and the plane of VCS). The real structure functions (SF) $W_i$ have a dynamical origin and depend on three kinematical variables: the momentum transfer $q^2(<0)$, the effective mass of the produced $\gamma p$ system, $W(>m)$, and the momentum transfer $t=(k-q)^2$. It is more convenient to use, instead of $t$, the equivalent variable $\cos\theta$, where $\theta$
is the production angle of the real photon (in the CMS of 
$\gamma^*+p\to \gamma +p$), with respect to the three-momentum of the virtual 
$\gamma^*$, with $-1\le \cos\theta\le 1$, so that  $W_i=W_i(W,q^2,\cos\theta)$. 

In order to calculate these SF's, let us define the electromagnetic current ${\cal J}_{\mu}$ for the process 
$\gamma^*+p\to \gamma +p$ as follows:
$${\cal M}(\gamma^*p\to p\gamma)=\epsilon_{\mu}{\cal J}_{\mu}(k,q,p_1,e),$$
where $\epsilon_{\mu}$ is the four-vector of the virtual photon polarization, so that $\epsilon\cdot q=0$, with the current conservation $q\cdot {\cal J}$=0. The SFs $W_i$ are determined by the following quadratic combinations of the components of this current:
$$ W_1=\displaystyle\frac{1}{2}Tr \overline{({\cal J}_x{\cal J}_x^* +{\cal J}_y{\cal J}_y^*)},$$
$$ W_2=\displaystyle\frac{1}{2}Tr \overline{({\cal J}_x{\cal J}_x^* -{\cal J}_y{\cal J}_y^*)},$$
$$ W_3=\displaystyle\frac{1}{2}Tr \overline{{\cal J}_0{\cal J}_0^*},$$
$$ W_4=\displaystyle\frac{1}{2}Tr\overline{({\cal J}_x{\cal J}_0^* -{\cal J}_0{\cal J}_y^*)},$$
where the line over the products of the current components denotes the sum over the real photon polarizations. These expressions are correct in the coordinate system (for the process $\gamma^*+p\to \gamma +p$) where the $z-$axis is along the three-momentum $\vec q$ and the $y-$axis is orthogonal to the reaction plane. 

It is important that the electromagnetic current ${\cal J}_{\mu}$ in the considered model is conserved, i.e. $q\cdot {\cal J}$=0, so we can replace the longitudinal component ${\cal J}_z$, by the time component ${\cal J}_0$, without violating the gauge invariance.

The calculation can be, therefore, performed analytically, using the following expressions for the current components:
$${\cal J}_x=A_1+i\vec\sigma\cdot\vec B_1,$$
$${\cal J}_y=A_2+i\vec\sigma\cdot\vec B_2,$$
$${\cal J}_0=A_3+i\vec\sigma\cdot\vec B_3,$$
with
$$A_1=(e_x\cos\theta-e_z\sin\theta)K_{\Delta},$$
$$A_2=e_y\cos\theta K_{\Delta},$$
\begin{equation} 
A_3=0,
\label{eq:eqjs}
\end{equation}
$$\vec B_1=e_y\left [\displaystyle\frac{\vec k}{ \omega}K_{\Delta}
+ \omega \vec Q(|\vec q |-q_0\cos\theta )K_{\pi}\right ],$$
$$\vec B_2=\displaystyle\frac{1}{2}\left (\vec e\sin\theta
-e_x\displaystyle\frac{\vec k}{ \omega}\right )K_{\Delta} -\omega
\vec Q\left [ e_x(|\vec q |-q_0\cos\theta )+e_zq_0\sin\theta
\right ] K_{\pi},$$
$$\vec B_3=-\vec Q\omega |\vec q |e_y K_{\pi} \sin\theta,$$
$$\vec Q=\displaystyle\frac{\vec q}{E_1+m}-\displaystyle\frac{\vec k}{E_2+m}, 
$$
where $K_{\pi}$ and $K_{\Delta}$ are:
$$K_{\pi}=\displaystyle\frac{g}{m_{\pi}}
\displaystyle\frac{F{\pi\gamma^* \gamma }(q^2)}
{t-m_{\pi}^2},$$
$$ K_{\Delta}=\displaystyle\frac{4\mu^2(\Delta\to N)G(q^2)M\omega|\vec q |}
{3(M+m)^2(W^2-M^2+iM\Gamma)}.$$
Using these formula one can find relatively simple analytical expressions for 
the SFs $W_i$:
$$ W_1=\displaystyle\frac{1}{4}(7+3\cos^2\theta)\left |K_{\Delta}\right |^2
+ \omega^2\left [(1+\cos^2\theta)(\vec {q^2}+q_0^2)-4\cos\theta q_0|\vec q |
\right ] K_{\pi}^2\Lambda, $$
$$+ \omega {\cal R}e K_{\pi}K_{\Delta}^*\left \{ 
-\displaystyle\frac{W-m}{W+m}
\left [|\vec q |(1+\cos^2\theta)-2q_0\cos\theta \right ] \right .
\left .+\displaystyle\frac{|\vec q |}{E_1+m}\left [-q_0(1+\cos^2\theta)+
2|\vec q |\cos\theta \right ] \right \},$$
$$W_2=\sin^2\theta\left [\displaystyle\frac{3}{4}\left |K_{\Delta}\right |^2
-\omega^2q^2K_{\pi}^2\Lambda \right .
\left .+ \omega|\vec q |\left(\displaystyle\frac{q_0}{E_1+m}-\displaystyle\frac{\omega}{E_2+m}\right )
{\cal R}e K_{\pi}K_{\Delta}^*\right],$$
$$W_3=\sin^2\theta\omega^2 |\vec q |^2 K_{\pi}^2 \Lambda,$$
$$W_4=-\sin\theta \left\{2\omega^2|\vec q |(|\vec q |-\cos\theta q_0)
K_{\pi}^2\Lambda\right .
\left . +\left [\omega (E_1-m)\cos\theta-|\vec q |(E_2-m)\right ]{\cal R}e K_{\pi}K_{\Delta}^*\right \},$$
with
$$\Lambda=\displaystyle\frac{E_1-m}{E_1+m}+\displaystyle\frac{E_2-m}{E_2+m}
-\displaystyle\frac{2\omega|\vec q |\cos\theta}{(E_1+m)(E_2+m)}.$$

As only the transversal virtual photon polarization in $\gamma^*+p\to \gamma +p$
is contributing for the $s-$channel $\Delta$-excitation, the different terms which appear in the SFs $W_i$ have a transparent meaning.  The numerical results are shown in Fig. \ref{fig:fig3} and \ref{fig:fig4} for $W_1$ to $W_4$ from top to bottom. The  columns correspond to $-q^2=1$, 3, 5 GeV$^2$, from left to right. The solid, dashed, dotted lines correspond to $W=1.1$, 1.232 , and 1.38 GeV.

The SFs $W_1$ and $W_2$ are positive and comparable in the considered kinematical region. 
The SF $W_3$, which describes the absorption of virtual photon with longitudinal polarization is driven only by one-pion exchange. The ratio $W_3/W_1$ is increasing with $-q^2$, in accordance with the different behavior of the  $\pi\gamma^*\gamma$ and $\gamma^*N\to \Delta$ form factors.

Fig. \ref{fig:fig5} illustrates the relative role of the different contributions to the SF $W_1$: in particular the one-pion exchange and the $\pi\times\Delta$-interference are important in the forward angular region.
\section{Polarization observables}
Let us derive the dependence of the exclusive cross section 
$\vec p(\vec e,
e\gamma)p$ on the target polarization $\vec{\cal T}$. In terms of the quantity $X_1$ defined above, we can write:
$$X_1 =\hat{\vec m}\cdot \vec{\cal T}\left ( \sqrt{1-\epsilon^2}T_1-\cos\phi 
\sqrt{2\epsilon(1-\epsilon)\displaystyle\frac{(-q)^2}{|\vec q |^2}}T_2
\right )$$
$$+ \hat{\vec q}\cdot \vec{\cal T}\left ( \sqrt{1-\epsilon^2}T_3-\cos\phi 
\sqrt{2\epsilon(1-\epsilon)\displaystyle\frac{(-q)^2}{|\vec q |^2}}T_4
\right )$$
$$+ \hat{\vec n}\cdot \vec{\cal T}
\sin\phi \sqrt{2\epsilon(1-\epsilon)\displaystyle\frac{(-q)^2}{|\vec q |^2}}T_5.
$$
The unit vectors, $\hat{\vec n}$, $\hat{\vec q}$ and $\hat{\vec m}$ are defined as follows:
$$\hat{\vec n}=\displaystyle\frac {\vec q\times \vec k}{|\vec q\times \vec k|},~\hat{\vec q}=\displaystyle\frac {\vec q}{|\vec q|},~\hat{\vec m}=
\hat{\vec n}\times\hat{\vec q}.$$
The real structure functions $T_1-T_5$, which characterize the T-even asymmetries for $\vec e\vec p$- interaction, can be calculated using the following expression:
$$T_1=\displaystyle\frac {i}{2}Tr\left (
\overline{
{\cal J}_x\vec\sigma\cdot\hat{\vec m} {\cal J}_y^*-
{\cal J}_y\vec\sigma\cdot\hat{\vec m} {\cal J}_x^*}\right ),$$
$$T_2=\displaystyle\frac {i}{2}Tr\left (
\overline{
{\cal J}_y\vec\sigma\cdot\hat{\vec m} {\cal J}_0^*-
{\cal J}_0\vec\sigma\cdot\hat{\vec m} {\cal J}_y^*}\right ),$$
$$T_3=\displaystyle\frac {i}{2}Tr\left (
\overline{
{\cal J}_x\vec\sigma\cdot\hat{\vec q} {\cal J}_y^*-
{\cal J}_y\vec\sigma\cdot\hat{\vec q} {\cal J}_x^*}\right ),$$
$$T_4=\displaystyle\frac {i}{2}Tr\left (
\overline{
{\cal J}_y\vec\sigma\cdot\hat{\vec q} {\cal J}_0^*-
{\cal J}_0\vec\sigma\cdot\hat{\vec q} {\cal J}_y^*}\right ),$$
$$T_5=\displaystyle\frac {i}{2}Tr\left (
\overline{
{\cal J}_x\vec\sigma\cdot\hat{\vec n} {\cal J}_0^*-
{\cal J}_0\vec\sigma\cdot\hat{\vec n} {\cal J}_x^*}\right ).$$
Using the expressions  (\ref{eq:eqjs}) for the current components, 
${\cal J}_x$, ${\cal J}_y$ and ${\cal J}_0$, one can find for the polarized structure functions $T_i$ the following analytical expressions, which are valid in the considered model:
$$T_1= \sin^2 \theta\left [\omega(E_1-m)-3\cos\theta|\vec q |(E_2-m)\right ]{\cal R}e K_{\pi}K_{\Delta}^*,$$
$$T_2= -\displaystyle\frac{3}{4}\sin\theta\cos\theta\left |K_{\Delta}\right |^2+
\displaystyle\frac{1}{2}\sin\theta\left \{ (\cos\theta q_0-|\vec q |)
\displaystyle\frac{\omega|\vec q |}{(E_1+m)}-\right .$$
$$
\left .
(E_2-m)\left [ -5\cos\theta|\vec q |
+(2+3\cos^2\theta)q_0\right ] \right \}{\cal R}e K_{\pi}K_{\Delta}^*,$$
$$T_3= \sin \theta\left [2\omega(E_1-m)\cos\theta+(1-3\cos^2\theta)|\vec q |
(E_2-m)\right ]{\cal R}e K_{\pi}K_{\Delta}^*,$$
$$T_4= -\displaystyle\frac{1}{4}(1+3\cos^2\theta)\left |K_{\Delta}\right |^2+
\displaystyle\frac{1}{2}\left \{ \displaystyle\frac{2\omega|\vec q |}{(E_1+m)}
\left [(1+\cos^2\theta)q_0-2\cos\theta|\vec q |\right ]-
\right .$$
$$
\left .
(E_2-m)
\left [q_0\cos\theta(1+3\cos^2\theta)+|\vec q |(1-5\cos^2\theta)
\right ]\right \}{\cal R}e K_{\pi}K_{\Delta}^*,$$
$$T_5= -\sin^2 \theta\omega(E_1-m){\cal R}e K_{\pi}K_{\Delta}^*.$$
Similar expressions can be found in case of unpolarized target and measurement of the polarization of the outgoing proton. Let's call $P_1$ to $P_5$ the corresponding polarized structure functions:

$$P_1= -\sin^2 \theta\left [\omega(E_1-m)+\cos\theta|\vec q |(E_2-m)\right ]{\cal R}e K_{\pi}K_{\Delta}^*,$$
$$P_2= -\displaystyle\frac{1}{4}\sin\theta\cos\theta\left |K_{\Delta}\right |^2+
\displaystyle\frac{1}{2}\sin\theta\left \{ (|\vec q |-\cos\theta q_0)
\displaystyle\frac{\omega|\vec q |}{(E_1+m)}+\right .$$
$$
\left .
(E_2-m)\left [ 3\cos\theta|\vec q |
+(2+\cos^2\theta)q_0\right ] \right \}{\cal R}e K_{\pi}K_{\Delta}^*,$$
$$P_3= \sin \theta\left [2\omega(E_1-m)\cos\theta-(1+\cos^2\theta)|\vec q |
(E_2-m)\right ]{\cal R}e K_{\pi}K_{\Delta}^*,$$
$$P_4= -\displaystyle\frac{1}{4}(3+\cos^2\theta)\left |K_{\Delta}\right |^2+
\left \{ \displaystyle\frac{\omega|\vec q |}{(E_1+m)}
\left [(1+\cos^2\theta)q_0-2\cos\theta|\vec q |\right ]+
\right .$$
$$
\left .
(E_2-m)\displaystyle\frac{1}{2}
\left [|\vec q |(1+3\cos^2\theta)-q_0\cos\theta(3+\cos^2\theta)
\right ]\right \}{\cal R}e K_{\pi}K_{\Delta}^*,$$
$$P_5= \sin^2 \theta\omega(E_1-m){\cal R}e K_{\pi}K_{\Delta}^*.$$

So the numerical calculations  for these polarized structure functions
$T_i(W,q^2,\cos\theta)$ and $P_i(W,q^2,\cos\theta)$ can be done in the same framework as for the calculation
of the unpolarized structure functions $W_i$. More precisely, we show in Figs. \ref{fig:fig5} and \ref{fig:fig6} the ratios ${\cal T}_1=T_1/W_1$, to 
${\cal T}_5=T_5/W_1$ (from top to bottom), which is more convenient for the analysis of polarization phenomena. Similarly, in Figs.
\ref{fig:fig7} and \ref{fig:fig8} we show the ratios 
${\cal P}_1=P_1/W_1$, to ${\cal P}_5=P_5/W_1$ (from top to bottom).

The notations are as in Fig. \ref{fig:fig3}. All these 'reduced' polarized SFs are comparable in absolute value, being in the limits $\pm 0.5$. They are almost independent on $q^2$, but depend strongly on W, in the considered interval. The $\theta$-dependence is driven by the relative role of the two mechanisms.
All polarized SFs show large sensitivity to the relative sign of the $\pi$ and $\Delta$-contributions (when $W\ne M$).

It is necessary to keep in mind, that these large polarization phenomena, for VCS, have to be multiplied, (after integration over the experimental acceptance)
by the electromagnetic constant $\alpha$, in order to have preliminary estimation of this effect on radiative corrections for elastic $ep$-scattering. 
We are aware that this is an oversimplified estimation, which can not be directly applied to the experimental observables in $ep$-elastic scattering, however it shows that a complete evaluation of radiative corrections to the precise $ep$-elastic scattering data of \cite{Jo00} have to include  corrections due to VCS (and to its interference with the Bethe-Heitler mechanism).

\section{Conclusions}
We calculated the T-even polarization observables for virtual Compton scattering,for two cases: $\vec e+\vec p\to e+p+\gamma$ and $\vec e+ p\to e+ \vec p+\gamma$ in particular kinematical conditions, at relatively large momentum transfer and for $p\gamma$ excitation energy from threshold to the $\Delta$ region. We used a realistic model for $\gamma^*+ p\to\gamma+ p$ which takes into account two main contributions: the $\Delta$-excitation in $s-$channel and the $\pi$-exchange in $t-$channel. The main advantage of this model is that the necessary parameters such as the strong interaction couplings and the electromagnetic form factors are well known. We considered polarized target effects as well as all components of the final proton polarization, in order to estimate the order of magnitude of polarization effects for VCS, considered as an independent interesting process. All calculated asymmetries and polarization components are large in absolute value (for $W\ne M$), showing a specific scaling behavior, i.e. the considered polarization phenomena are nearly $q^2$-independent. The polarized SFs are sensitive to the relative sign of two fundamental constants, $g$ and $F_{\pi}$. In this respect data from VCS at large momentum transfer could determine, in principle, this sign. 

We calculated all four unpolarized structure functions and shown the importance of the contributions due to the absorbtion of longitudinally polarized photons, through $\pi$-exchange.
 
The extension to other polarization observables is straightforward. 

These results can be applied for the estimation of the corresponding contributions to radiative corrections to polarization phenomena in $ep$-elastic scattering.

\newpage
\begin{figure}
\begin{center}
\mbox{\epsfxsize=14.cm\leavevmode \epsffile{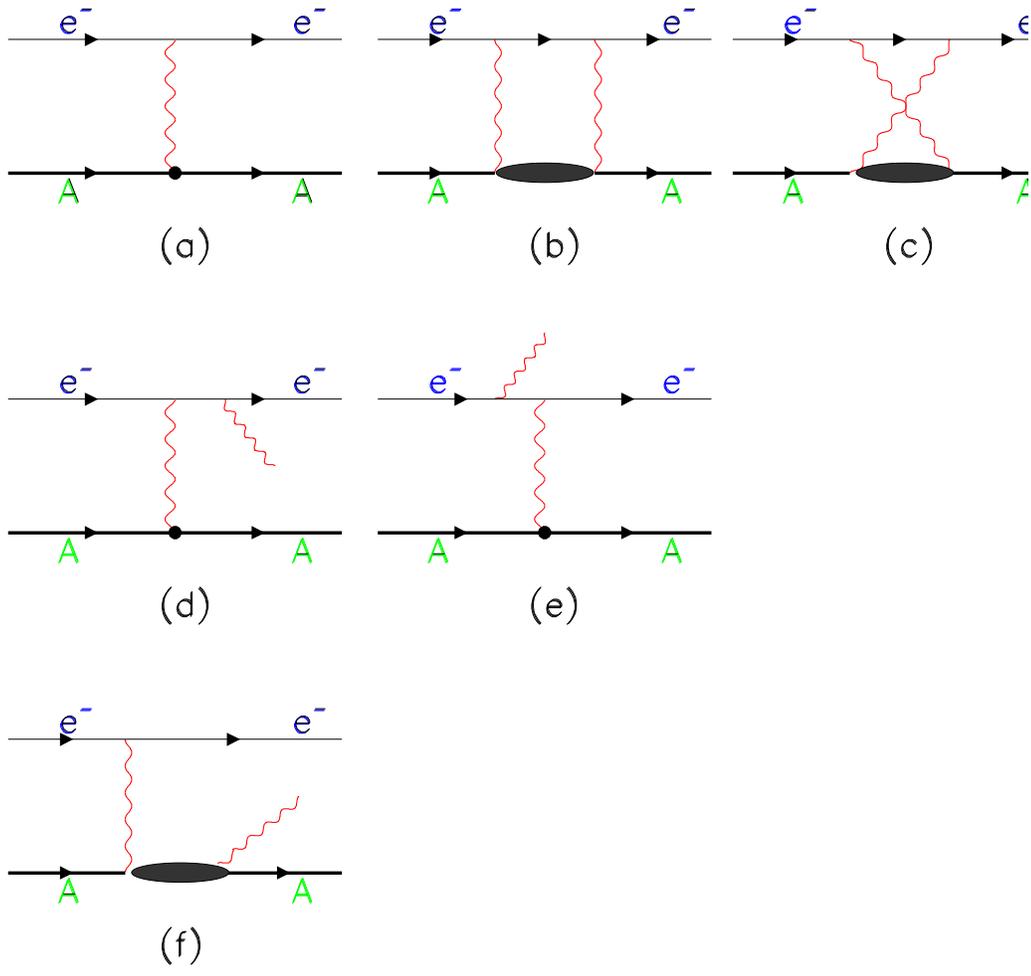}}
\end{center}
\caption{Relevant Feynman diagrams in the calculation of radiative corrections for $ep$-cattering: one-photon exchange (a); two-photon exchange
(b-c);
Bethe-Heitler mechanism (d-e); virtual Compton scattering (f).}
\label{fig:fig1}
\end{figure}
\begin{figure}
\begin{center}
\mbox{\epsfxsize=14.cm\leavevmode \epsffile{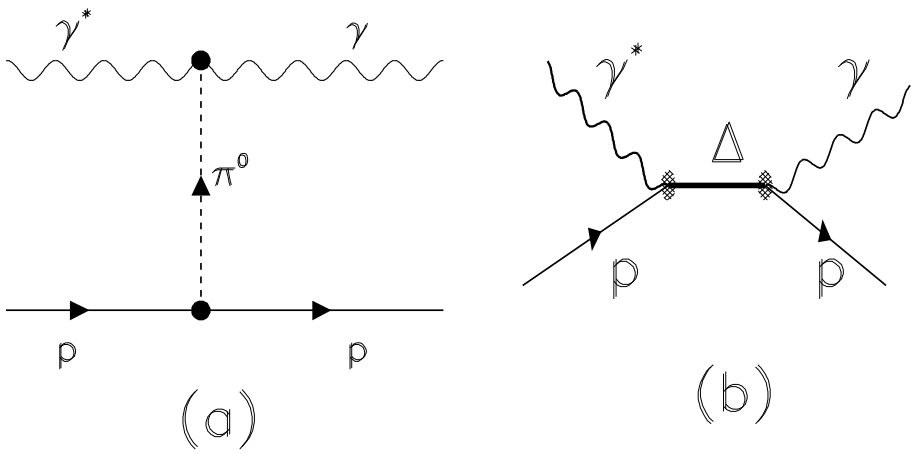}}
\end{center}
\caption{ Feynman diagrams for $\gamma^*+ p\to\gamma+ p$: one-pion exchange in $t$-channel (a); 
$\Delta$-echange in $s$-channel (b).}
\label{fig:fig2}
\end{figure}
\begin{figure}
\begin{center}
\mbox{\epsfxsize=14.cm\leavevmode \epsffile{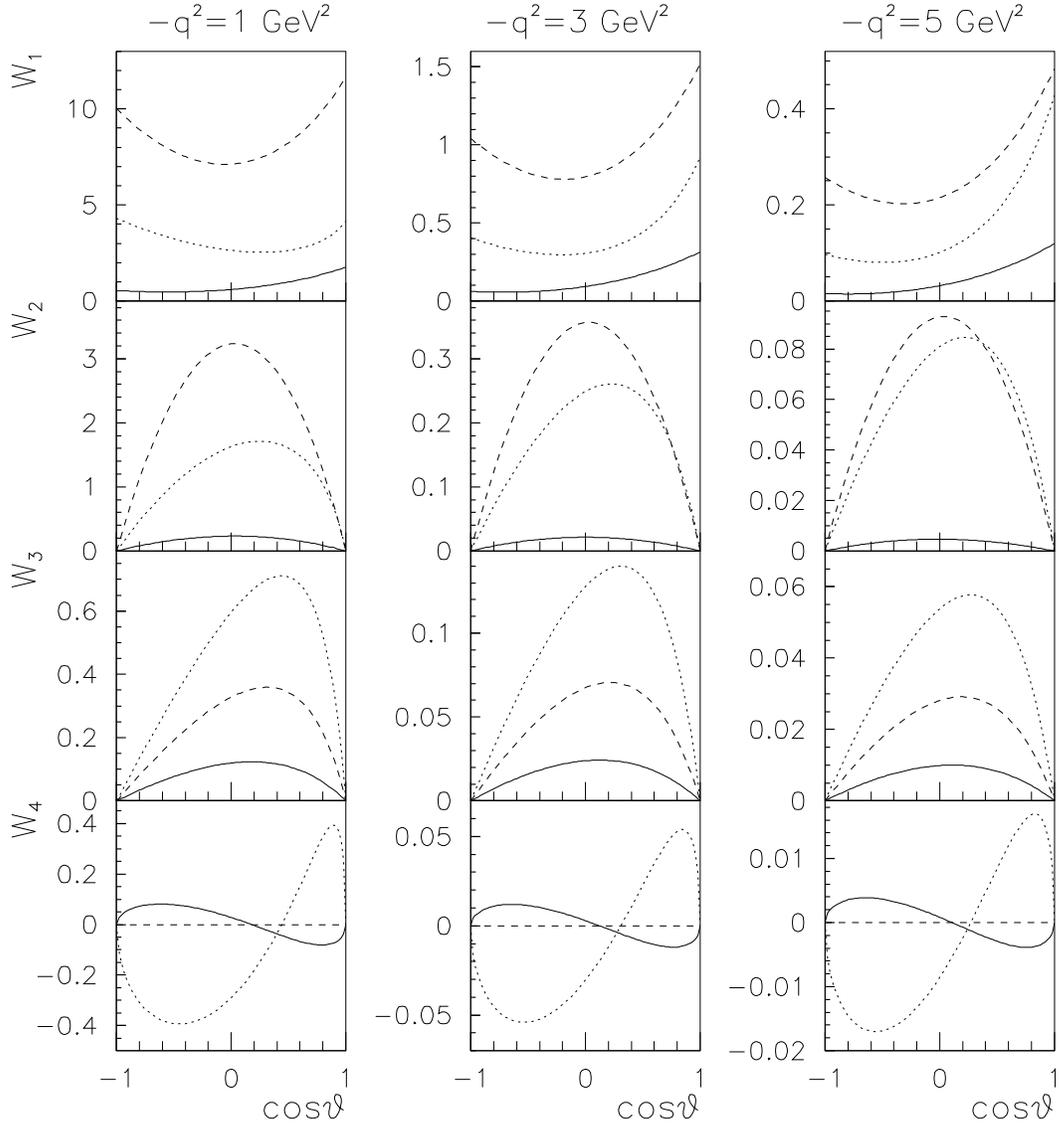}}
\end{center}
\caption{Structure functions $W_1$ to $W_4$ from top to bottom as functions of $\cos\theta$. The  columns correspond to $-q^2=1$, 3, 5 GeV$^2$, from left to right. The solid, dashed, dotted line correspond to $W=1.1$, 1.232, and 1.38 GeV. The calculation is done for the positive relative sign of $\Delta$ and $\pi$-contributions.}
\label{fig:fig3}
\end{figure}
\begin{figure}
\begin{center}
\mbox{\epsfxsize=14.cm\leavevmode \epsffile{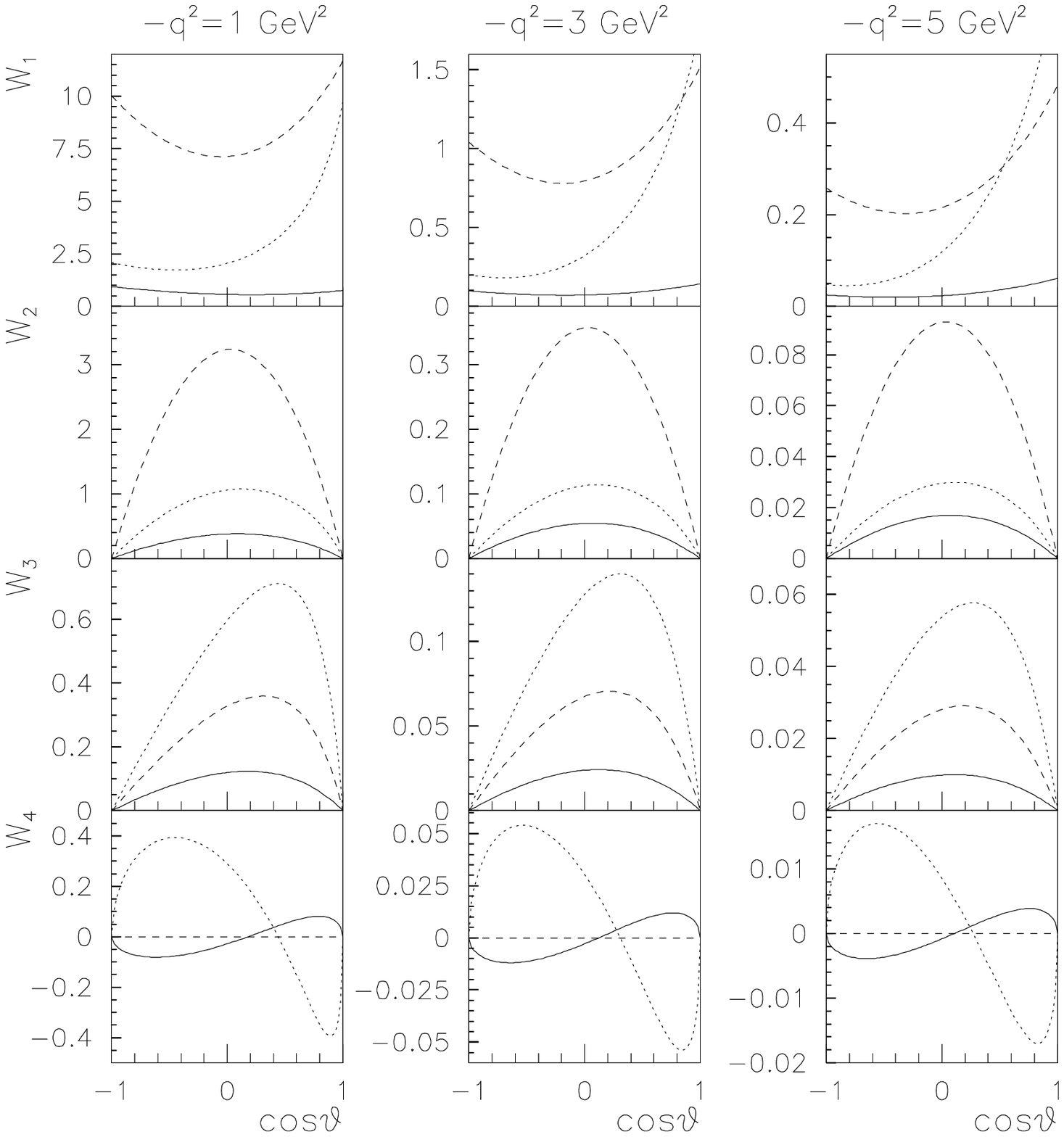}}
\end{center}
\caption{Same as Fig. 3, for negative relative sign of $\Delta$ and $\pi$-contributions.}
\label{fig:fig4}
\end{figure}
\begin{figure}
\begin{center}
\mbox{\epsfxsize=14.cm\leavevmode \epsffile{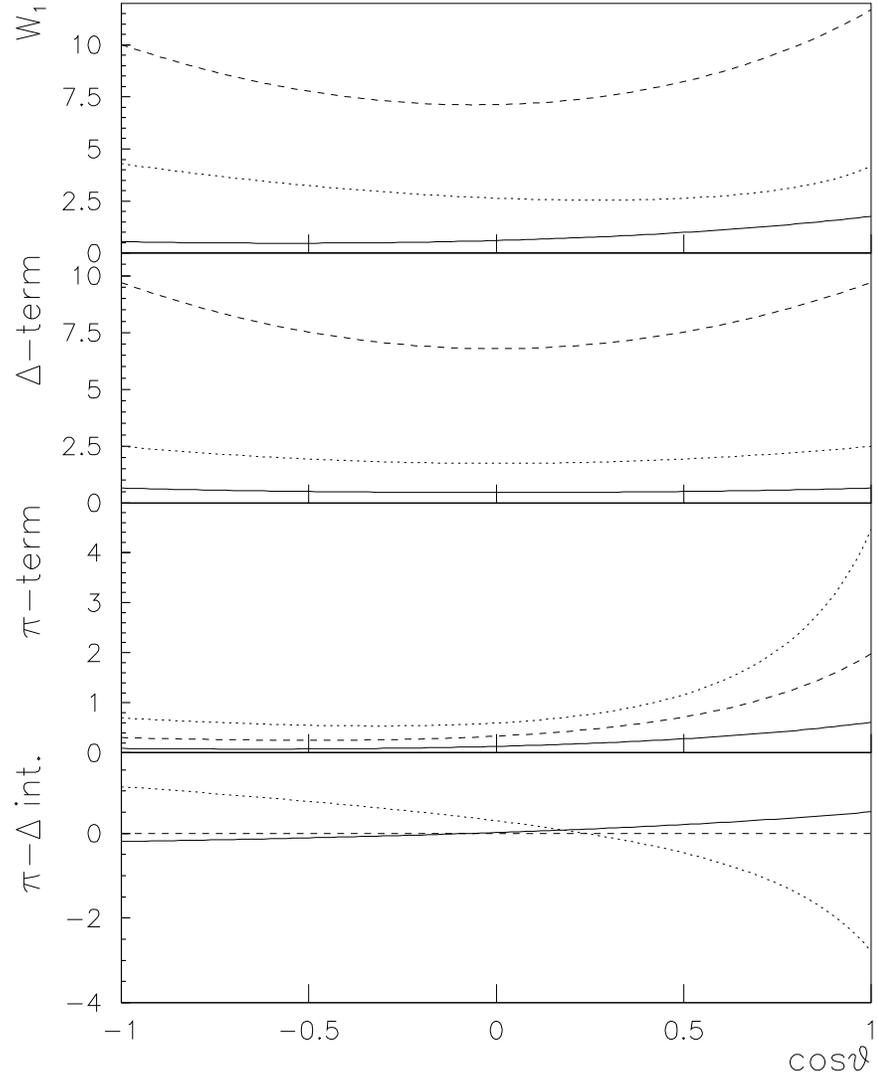}}
\end{center}
\caption{From top to bottom: angular dependence of $W_1$, and of the different contributions to $W_1$: $\Delta$, $\pi$ and $\Delta-\pi$-interference 
for positive relative sign of $\Delta$ and $\pi$-contributions. Notations as in Fig. 3.}
\label{fig:fig5}
\end{figure}
\begin{figure}
\begin{center}
\mbox{\epsfxsize=14.cm\leavevmode \epsffile{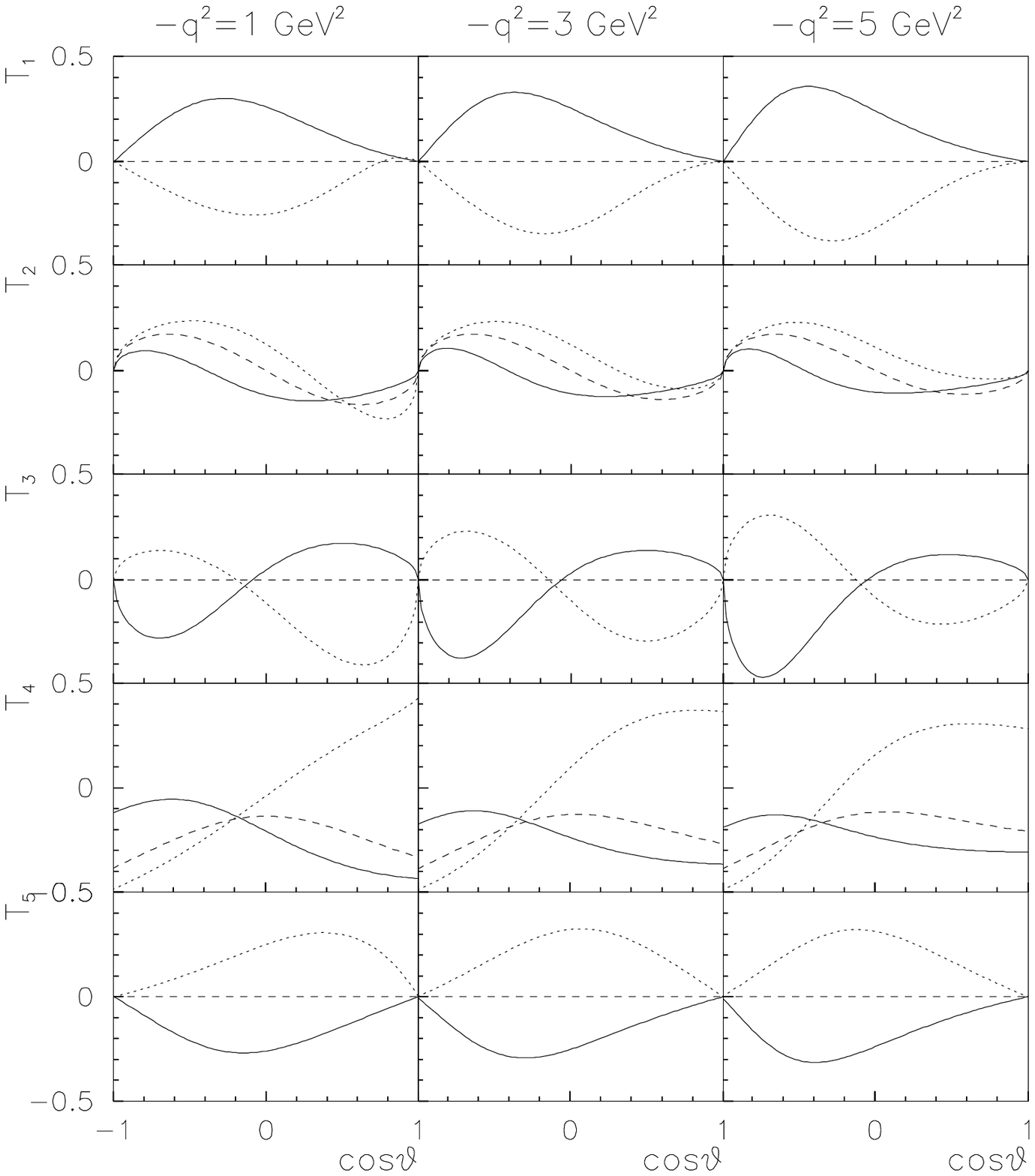}}
\end{center}
\caption{Polarized 'reduced' structure functions $T_1/W_1$ to $T_5/W_1$ 
from top to bottom as functions of $\cos\theta$. Notations as in Fig. 3. The calculation is done for positive relative sign of $\Delta$ and $\pi$-contributions.}
\label{fig:fig6}
\end{figure}
\begin{figure}
\begin{center}
\mbox{\epsfxsize=14.cm\leavevmode \epsffile{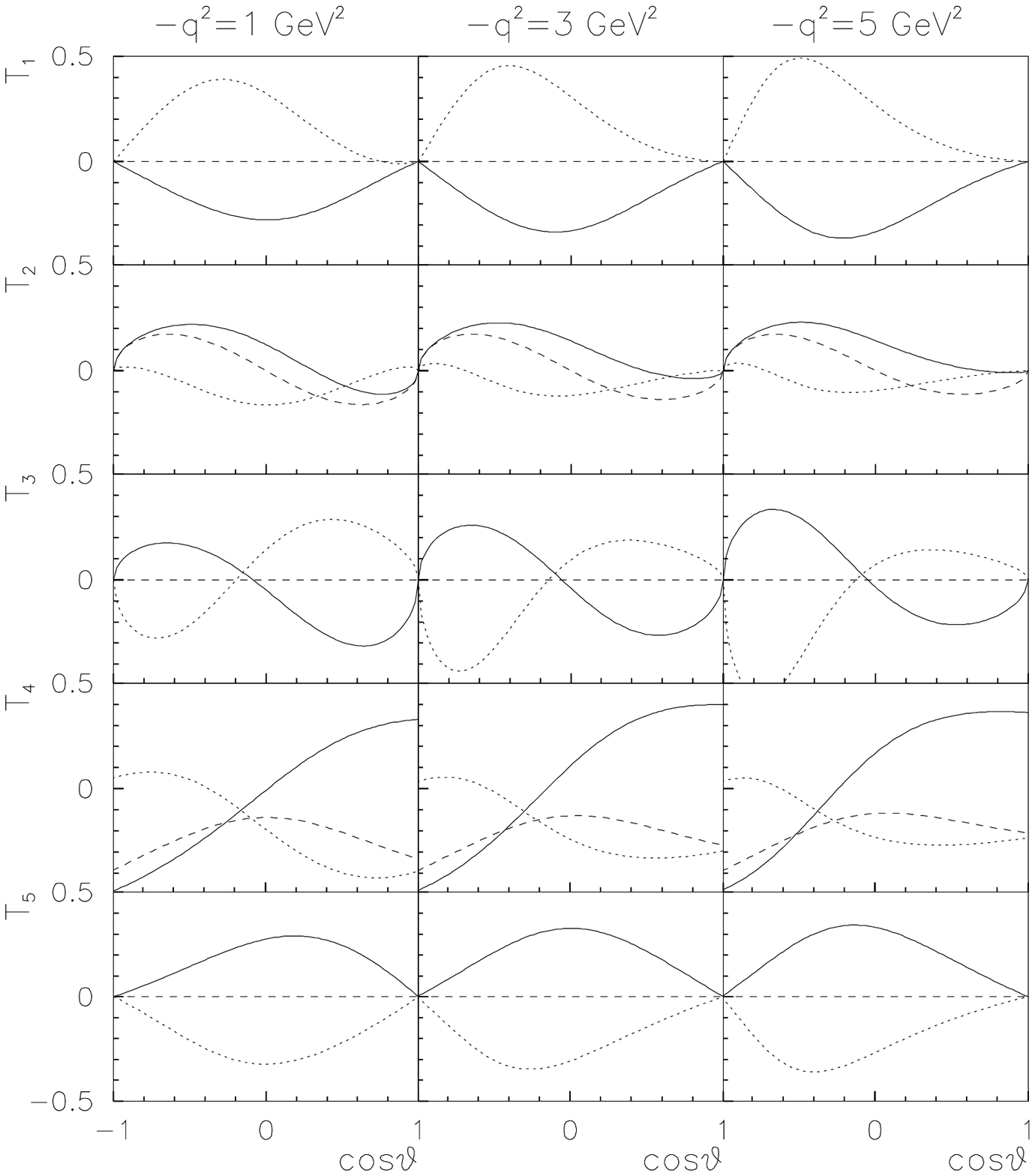}}
\end{center}
\caption{Same as Fig. 6, but  for negative relative sign of $\Delta$ and $\pi$-contributions.}
\label{fig:fig7}
\end{figure}
\begin{figure}
\begin{center}
\mbox{\epsfxsize=14.cm\leavevmode \epsffile{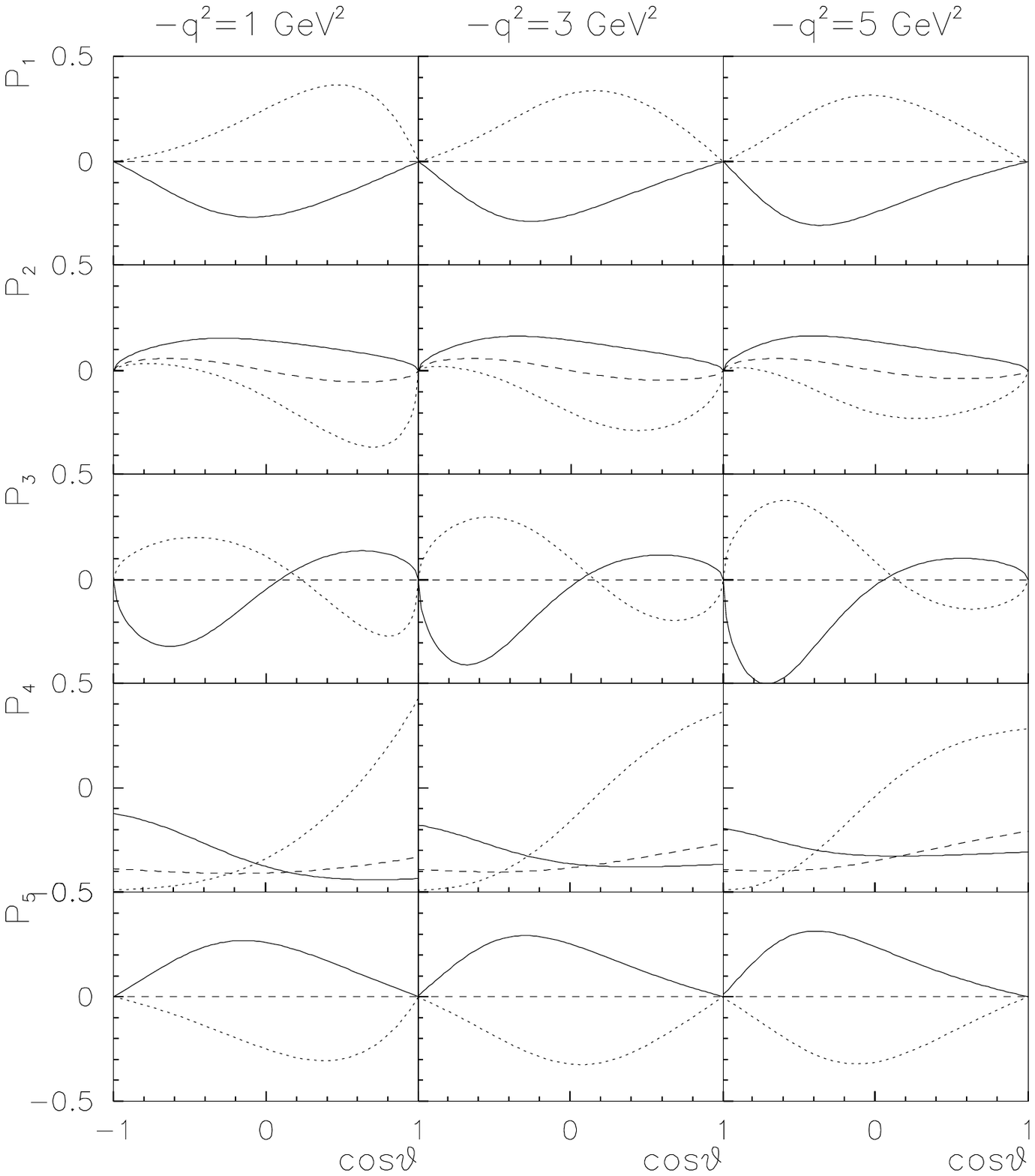}}
\end{center}
\caption{Polarized 'reduced' structure functions $P_1/W_1$ to $P_5/W_1$ 
from top to bottom as functions of $\cos\theta$. Notations as in Fig. 3. 
The calculation is done for positive relative sign of $\Delta$ and $\pi$-contributions.}
\label{fig:fig8}
\end{figure}
\begin{figure}
\begin{center}
\mbox{\epsfxsize=14.cm\leavevmode \epsffile{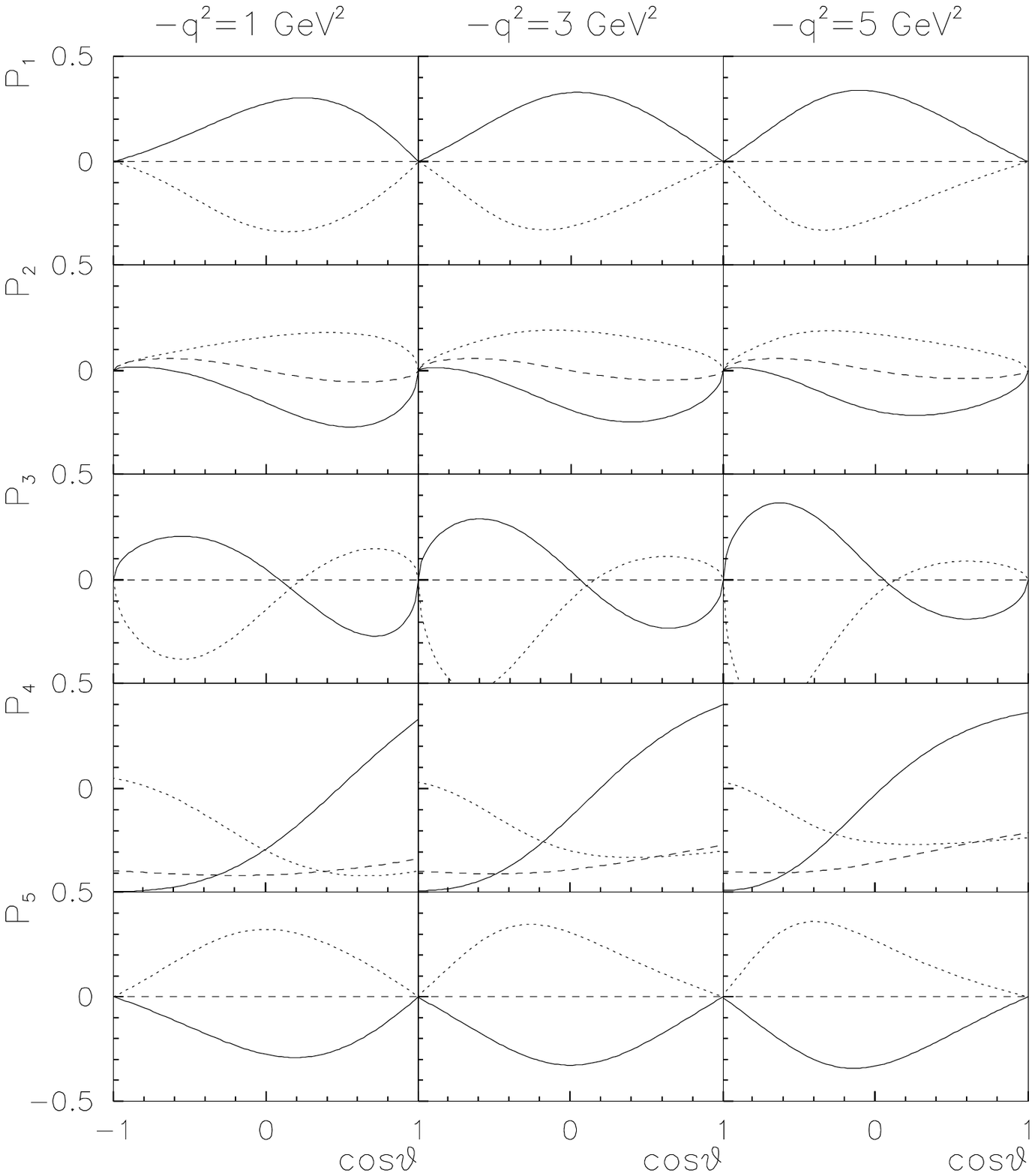}}
\end{center}
\caption{Same as Fig. 8, but for negative relative sign of $\Delta$ and $\pi$-contributions.}
\label{fig:fig9}
\end{figure}
\end{document}